\documentclass[12pt]{iopart}
\usepackage{graphicx}
\usepackage{amssymb}
\usepackage{esdiff}
\usepackage{subfig}
\usepackage{comment}
\usepackage{parskip}
\usepackage{hhline}
\usepackage{array}
\usepackage{textcomp}
\usepackage{hyphsubst}
\usepackage{hyperref}
\usepackage{blindtext}
\usepackage{siunitx}
\usepackage[dvipsnames]{xcolor}

\begin{document}

\title[Anodization-free fabrication process for cross-type Josephson tunnel junctions]{Anodization-free fabrication process for high-quality cross-type Josephson tunnel junctions based on a Nb/Al-AlO$_x$/Nb trilayer}

\author{F Adam$^{1,2}$, C Enss$^{1,3}$ and S Kempf$^{2,3,1}$}
\address{$^1$ Kirchhoff-Institute for Physics, Heidelberg University, Im Neuenheimer Feld 227, D-69120 Heidelberg, Germany}
\address{$^2$ Institute of Micro- and Nanoelectronic Systems, Karlsruhe Institute of Technology, Hertzstrasse 16, Building 06.41, D-76187 Karlsruhe, Germany}
\address{$^3$ Institute for Data Processing and Electronics, Karlsruhe Institute of Technology, Hermann-von-Helmholtz-Platz 1, Building 242, D-76344 Eggenstein-Leopoldshafen, Germany}
\ead{\mailto{fabienne.adam@kit.edu}}
\vspace{10pt}
\begin{indented}
\item[]March 2024
\end{indented}

% --------------------------------------------

\begin{abstract}
Josephson tunnel junctions form the basis for various superconductor electronic devices. For this reason, enormous efforts are routinely taken to establish and later on maintain a scalable and reproducible wafer-scale manufacturing process for high-quality Josephson junctions. Here, we present an anodization-free fabrication process for Nb/Al-AlO$_x$/Nb cross-type Josephson junctions that requires only a small number of process steps and that is in general intrinsically compatible with wafer-scale fabrication. We show that the fabricated junctions are of very high quality and, compared to other junction types, exhibit not only a significantly reduced capacitance but also an almost rectangular critical current density profile. Our process hence enables the usage of low capacitance Josephson junctions for superconductor electronic devices such as ultra-low noise dc-SQUIDs, microwave SQUID multiplexers based on non-hysteretic rf-SQUIDs and RFSQ circuits.
\end{abstract}

% --------------------------------------------

\vspace{2pc}

\maketitle

% --------------------------------------------

\section{Introduction}\label{sec:1}

Josephson tunnel junctions are key components of any superconductor electronic devices. This includes superconducting quantum bits \cite{Clarke2008}, superconducting quantum interference devices (SQUIDs) \cite{Fagaly2006}, rapid single flux quantum (RSFQ) circuits \cite{Yohannes2015,Kunert2017}, Josephson voltage standards \cite{Hamilton2000}, single electron transistors (SETs) \cite{Bluethner1996,Dolata2005}, Josephson parametric amplifiers \cite{Bhat1999,Aumentado2020} or superconductor-insulator-superconductor (SIS) mixers \cite{Rothermel1994,Westig2012}. Most of these devices are based on refractory Josephson tunnel junctions made of an {\it in-situ} deposited Nb/Al-AlO$_x$/Nb trilayer, the latter being an excellent choice regarding junction quality, tunability of the critical current density, scalability and run-to-run reproducibility of characteristic junction parameters as well as resilience to thermal cycling. A key requirement for realizing integrated circuits based on these junctions is the availability of a wafer-scale fabrication process \cite{Tolpygo2019,Wan2021,Chen2023}. For this reason, research facilities make huge efforts to establish and maintain a fabrication process for high-quality Nb/Al-AlO$_x$/Nb Josephson tunnel junctions. In some cases, these efforts are further challenged by the need for minimizing the junction capacitance $C_\mathrm{JJ}$ to allow, for example, improving the energy resolution of SQUIDs \cite{Clarke2004,Schmelz2011}.

The capacitance $C_\mathrm{JJ} = C_\mathrm{int} + C_\mathrm{par}$ of a Josephson tunnel junction is composed of an intrinsic and a parasitic contribution. The intrinsic capacitance $C_\mathrm{int}$ depends on the material and the dimensions of the tunnel barrier and is determined by barrier thickness $d$ (setting the critical current density) and the junction area $A_\mathrm{JJ}$. It scales inversely with the tunnel barrier thickness $d$. At the same time, the critical current density $j_\mathrm{c}$ scales exponentially with the tunnel barrier thickness $d$. For this reason, reducing the junction area $A_\mathrm{JJ}$ and simultaneously increasing the critical current density $j_\mathrm{c}$ effectively lowers the intrinsic junction capacitance assuming a fixed target value of the critical current $I_\mathrm{c}$. The parasitic capacitance $C_\mathrm{par}$ is due to overlaps of the superconducting wiring with the junction electrodes that are separated by the wiring insulation. It strongly depends on the fabrication technology, i.e. the type and thickness of insulation layers, the required actual overlap between wiring layers, etc..

In the past, several fabrication processes for Nb/Al-AlO$_x$/Nb Josephson tunnel junctions have been developed. These are based on reactive ion etching and wet-chemical anodization \cite{Gurvitch1983,Dolata1999}, chemical-mechanical polishing \cite{Ketchen1991,Bao1995}, focused ion beam etching \cite{Watanabe2004} or shadow evaporation \cite{Harada1994}. Though these processes are used with great success, they either yield junctions with high capacitance or are prone to process variations due to barrier inhomogeneities or lithographic misalignments. Moreover, wet-chemical anodization requires a galvanic connection to ground, necessitating a temporary electrical connection of electrically floating devices such as rf-SQUIDs or qubits to their environment which must be removed in later fabrication steps. This complicates the fabrication process and introduces potential steps for junction damage.

Within this context, we present a variant of a fabrication process for cross-type Josephson tunnel junctions \cite{Dolata1999,Dang1991,Anders2009} that does not depend on wet-chemical anodization. Our process is hence particularly suited for fabricating electrically floating superconducting quantum devices. At the same time, the junction capacitance is minimized. Moreover, our process requires only a small number of fabrication steps, has in general the potential for wafer-scale fabrication and yields junctions with very high tunnel barrier homogeneity.

% --------------------------------------------

\section{Description of fabrication process}\label{sec:2}

\begin{figure}
\centering
    \includegraphics[width=0.7\columnwidth]{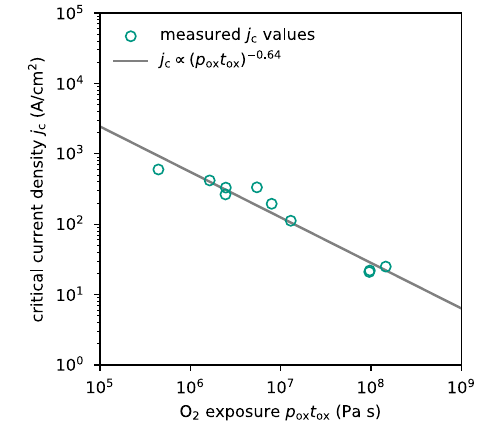}
    \caption{Measured dependence of the critical current density $j_{\mathrm{c}}$ on the oxygen exposure $p_{\mathrm{ox}}t_{\mathrm{ox}}$. The solid line is a fit to the measured data indicating the exponential dependence of the critical current density on the oxygen exposure.}
    \label{fig1}
\end{figure}

Our Josephson tunnel junctions are based on a Nb/Al-AlO$_x$/Nb trilayer that is \textit{in-situ} sputter-deposited on a thermally oxidized Si substrate. The thickness of the lower Nb base electrode, the Al layer and the upper Nb counter electrode are \SI{100}{\nano\meter}, \SI{7}{\nano\meter}, and \SI{100}{\nano\meter}, respectively. All layers are dc magnetron sputtered from 3\,inch targets in a high vacuum (HV) sputter system with a base pressure in the order of \SI{e-6}{\pascal}. During sputtering, the substrate is passively cooled by a thin layer of vacuum grease between substrate and sample holder. Prior to metal deposition, the substrate is pre-cleaned by an rf driven Ar plasma in the load-lock of the sputtering system. Both Nb layers are sputtered with a rate of \SI{0.63}{\nano\meter/\second} at a constant dc power of \SI{300}{\watt}. The pressure of the Ar atmosphere during sputtering is \SI{0.96}{\pascal} to yield Nb free of mechanical stress \cite{Kaiser2011}. The Al film is deposited in an Ar atmosphere with a pressure of \SI{0.72}{\pascal} using a dc power of \SI{100}{\watt} resulting in a deposition rate of \SI{0.31}{\nano\meter/\second}. For tunnel barrier formation within the load-lock of the sputtering system, the Al layer is oxidized at room temperature in a static O$_2$ atmosphere with pressure $p_\mathrm{ox}$. The critical current density $j_{\mathrm{c}}$ of the tunnel junctions depends on the total oxygen exposure $p_{\mathrm{ox}}t_{\mathrm{ox}}$ according to $j_{\mathrm{c}}\propto (p_{\mathrm{ox}}t_{\mathrm{ox}})^{-0.64}$ (see figure~\ref{fig1}). We typically vary the oxidation time $t_{\mathrm{ox}}$ at a fixed value of the oxidation pressure of $p_{\mathrm{ox}}= \SI{4}{\kilo\pascal}$.

\begin{figure}
\centering
    \includegraphics[width=0.7\columnwidth]{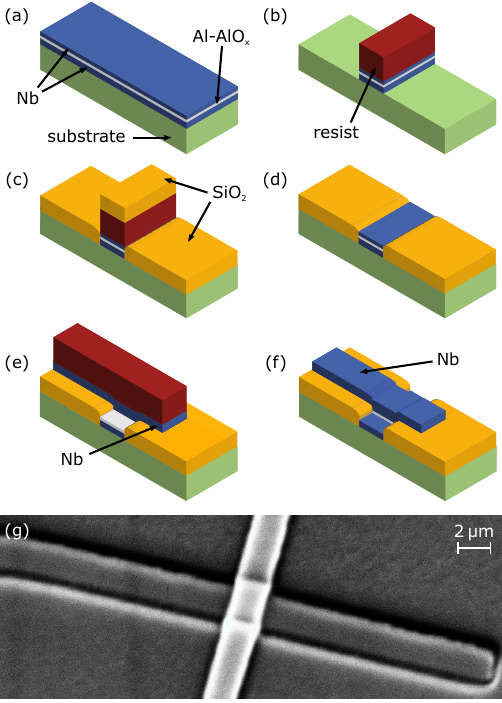}
    \caption{(a)-(f) Schematic overview of the different steps of our fabrication process for cross-type Josephson junctions. Shown are the state of the junction after (a) deposition of the Nb/Al-AlO$_x$/Nb trilayer, (b) trilayer patterning as a stripe, (c) deposition of the dielectric insulation layer for planarization, (d) removal of the photoresist mask, (e) deposition and patterning of the Nb wiring layer, and (f) removal of the residual Al and photoresist. Dimensions are not to scale. (g) Scanning electron microscope image of a finished cross-type Josephson junction.}
    \label{fig2}
\end{figure}

Figure~\ref{fig2} shows the individual fabrication steps for our cross-type Josephson junctions. After trilayer deposition (see figure~\ref{fig2}(a)), a positive, high-resolution UV photoresist (AZ MIR 701 29CP supplied by Microchemicals GmbH) is spin-coated on top of the trilayer and patterned as a narrow stripe using direct laser lithography. The width of this stripe defines one of the lateral dimensions of the final Josephson junction (see below). The resulting photoresist mask is used for etching the entire trilayer stack (see figure~\ref{fig2}(b)). Both Nb layers are etched by inductively coupled plasma reactive ion etching (ICP-RIE) using SF$_6$ and Ar in a mixing ratio of 2:1 at a constant pressure of \SI{2}{\pascal} as process gas. The rf power and the ICP power are \SI{10}{\watt} and \SI{300}{\watt}, respectively, resulting in an etch rate of \SI{2.5}{\nano\meter/\second}. The Al-AlO$_x$ layer and the thermal oxide of the Si substrate, respectively, act as etch stop for the ICP-RIE processes. The Al-AlO$_x$ layer is wet-chemically etched with an etching solution consisting of phosphoric acid, nitric acid, acetic acid and water that are mixed in a ratio of $16\negmedspace :\negmedspace 1\negmedspace :\negmedspace 1\negmedspace :\negmedspace 2$. As will be shown in section~\ref{sec:4}, wet-chemical etching of the Al layer is key to guarantee a high junction quality when omitting wet-chemical anodization.

The next step is the deposition of a dielectric insulation layer (see figure~\ref{fig2}(c)). The insulation is intended not only to protect the sidewalls of the patterned trilayer stripe, but also to quasi-planarize the trilayer stack before the deposition of subsequent layers. For insulator deposition, we use the same photoresist mask as for trilayer patterning, i.e. the mask is not removed after the prior etching steps. It is important to note that the photoresist acts as a shadow mask during dc magnetron sputter deposition, resulting in trenches beside the trilayer stripe as can be seen in the scanning electron microscope (SEM) image of a finished junction in figure~\ref{fig2} (g). We empirically found that the thickness of the insulation layer at the lowest point of these trenches is only about $50\,\%$ of the nominally deposited material. For this reason, the thickness of the insulation layer must be at least twice the layer thickness of the Nb base electrode to prevent shorts between the base electrode and subsequent wiring layers. We hence deposit a \SI{220}{\nano\meter} thick SiO$_2$ layer by rf magnetron sputtering utilizing a separate HV sputtering system, a gas mixture consisting of $60\,\%$ Ar and $40\,\%$ O$_2$ at a constant pressure of \SI{0.7}{\pascal} as process gas, and an rf power of \SI{250}{\watt}. This results in an overall deposition rate of \SI{1.3}{\nano\meter/\second}.

After removal of the photoresist mask (see figure~\ref{fig2}(d)), a Nb wiring layer with a thickness of \SI{200}{\nano\meter} is dc magnetron sputter deposited using a HV sputter system with a base pressure below \SI{6e-6}{\pascal} and a 2\,inch Nb target. The Ar pressure and the dc power are \SI{0.3}{\pascal} and \SI{70}{\watt}, respectively, resulting in a deposition rate of \SI{0.3}{\nano\meter/\second}. Prior to the deposition, the surface of the Nb counter electrode (upper Nb layer of the trilayer stripe) is pre-cleaned by Ar ion milling to remove native oxides and hence to ensure a superconducting contact between the counter electrode and the deposited Nb layer. This final layer is patterned by structuring a high-resolution UV photoresist (same as for trilayer patterning) as a narrow stripe that is oriented perpendicular to the embedded trilayer stripe and ICP-RIE for Nb etching. The top Nb layer of the trilayer stack is etched within the same etching cycle to define the final size of the counter electrode (see figure~\ref{fig2}(e)). By this, we yield rectangular Josephson tunnel junctions from the overlap of the trilayer and Nb wiring stripe. Finally, the residual Al-AlO$_x$ of the trilayer is removed by wet etching to enable later electrical contacts to the Nb base electrode (see figure~\ref{fig2}(f)).

It is worth mentioning that the area of our cross-type Josephson junctions is only limited by the resolution of the lithographic tool and not by alignment accuracy. Due to the minimum structure size of our laser lithography tool of \SI{1}{\micro\meter}, we are able to reliably fabricate cross-type junctions with a nominal area of \qtyproduct{1 x 1}{\micro\meter}, but even sub-micrometer-sized junctions are achievable with the help of e.g. DUV steppers or electron beam lithography. Even though such small junctions require higher values of the critical current density to achieve a target value of the critical current, the total junction capacitance is reduced as the intrinsic capacitance $C_{\mathrm{int}}$ linearly decreases with the junction area $A$ while the intrinsic capacitance per unit area $C_{\mathrm{int}}'$ only logarithmically increases with the critical current density $j_{\mathrm{c}}$ \cite{Maezawa1995,Tolpygo2007}. In addition, the capacitance of cross-type junctions has a negligible parasitic contribution as there are no direct wiring overlaps. Besides that, only two lithographic layers are required during the entire fabrication process. The higher values of the critical current density further lower the time taken to fabricate a batch of cross-type junctions as, according to figure~\ref{fig1}, the oxidation time for the formation of the tunnel barrier gets significantly shorter assuming a fixed oxidation pressure. 

% --------------------------------------------

\section{Experimental techniques for junction characterization}\label{sec:3}

\begin{figure}
\centering
    \includegraphics[width=0.7\columnwidth]{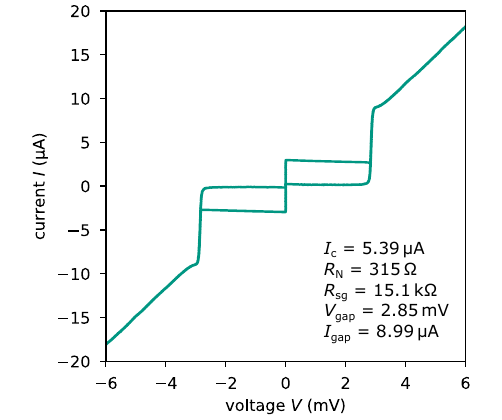}
    \caption{Current-voltage characteristic of one of our cross-type junctions with a target area of \qtyproduct{1 x 1}{\micro\meter} recorded at a temperature of $T= \SI{4.2}{\kelvin}$. All figures of merit except for the critical current $I_{\mathrm{c}}$ were taken directly from the characteristic. The critical current was obtained by the method described in section~\ref{sec:3}.}
    \label{fig2-3}
\end{figure}

Up to now, we have successfully fabricated more than 15 batches of cross-type junctions with linear dimensions varying between \SI{1.0}{\micro\meter} and \SI{4.2}{\micro\meter} using our anodization-free fabrication process. The characteristic figures of merit and hence the quality of fabricated junctions as well as their uniformity across an entire wafer were determined by recording the current-voltage ($IV$) characteristics (see figure~\ref{fig2-3} as an example)  of a sub-sample of each batch at a temperature of $T= \SI{4.2}{\kelvin}$ in a differential four-wire configuration. The utilized measurement set-up comprises low-pass filters at room and cryogenic temperatures to filter external rf interference signals. The dc bias current $I$ is generated by applying a triangular voltage signal $V_{\mathrm{gen}}$ with a frequency of \SI{3}{\hertz} to the series connection of all resistors in the input circuit of the set-up. This includes the equivalent resistance $R_{\mathrm{LPF}} = \SI{10.4}{\kilo\ohm}$ of both rf filters as well as the voltage-dependent resistance $R(V)=V/I$ of the Josephson junction to be measured. The actual bias current through the junction hence depends on the voltage drop $V$ across the junction and is given by
\begin{equation}\label{eq1}
    I=\frac{V_{\mathrm{gen}}}{R_{\mathrm{LPF}}}\left(1-\frac{V}{V_{\mathrm{gen}}}\right)\,.
\end{equation}
The voltage drop $V$ is measured using a battery-powered differential amplifier. To screen the samples from disturbances induced by variations of magnetic background fields, the cryo-probe is equipped with a mu-metal and a superconducting shield made of Nb.

The quality of each Josephson junction is evaluated by means of different figures of merit, among those the critical current $I_\mathrm{c}$, the normal state resistance $R_\mathrm{N}$, the gap voltage $V_{\mathrm{gap}}$, and the subgap resistance $R_\mathrm{sg}$. As conventional, the subgap resistance is determined at a voltage value of $V = \SI{2}{\milli\volt}$ in the subgap region. From these quantities, the characteristic resistance ratio $R_\mathrm{sg}/R_\mathrm{N}$ and the $I_{\mathrm{c}}R_{\mathrm{N}}$ product are calculated. The former is a junction area independent figure of merit to quantify subgap leakage, e.g. due to defects or shorts in the tunnel barrier \cite{Kuroda1988, Du2007}. The latter is a measure for the strength of Cooper pair tunneling that depends on the profile of the electric potential along the tunnel barrier \cite{Likharev1986}.

\begin{figure}
\centering
    \includegraphics[width=0.7\columnwidth]{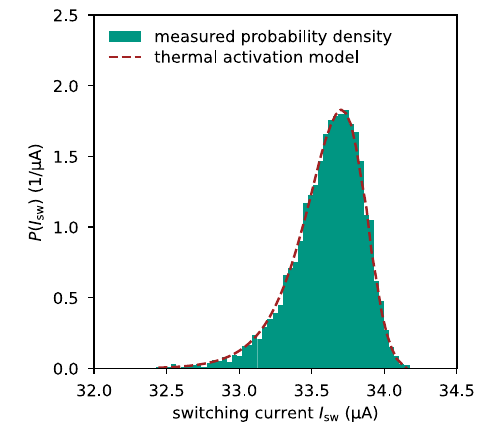}
    \caption{Switching current distribution $P(I_{\mathrm{sw}})$ of a \qtyproduct{3.4 x 3.4}{\micro\meter}-sized cross-type junction recorded at a temperature of $T=\SI{4.2}{\kelvin}$. For the measurement, the switching current of the junction was measured 5000 times by ramping up the bias current and recording the current values at which the junction switches from the superconducting into the normal conducting state. The solid red line represents a fit for a critical current of $I_{\mathrm{c}}=\SI{38.6}{\micro\ampere}$ according to the thermal activation model represented by equations~\ref{eq2} and \ref{eq3}.}
    \label{fig3}
\end{figure}

The critical current $I_\mathrm{c}$ of each Josephson junction was determined by its relation $I_\mathrm{c} = \kappa I_\mathrm{gap}$ \cite{Likharev1979} to the gap current $I_{\mathrm{gap}}$ as the switching current $I_\mathrm{sw}$ at $T~=~\SI{4.2}{\kelvin}$, extracted from $IV$-characteristics, is always significantly smaller than the true critical current. The gap current corresponds to the amplitude of the quasiparticles' tunneling current at the gap voltage. The deviation between the switching current $I_{\mathrm{sw}}$ and the critical current $I_\mathrm{c}$ is caused by thermal noise \cite{Fulton1974, Danchi1984}. The recursion formula
\begin{equation}\label{eq2}
    P(I_{\mathrm{sw}})=\tau^{-1}(I_\mathrm{sw}) \left(\frac{\mathrm{d}I}{\mathrm{d}t}\right)^{-1}\left[1-\int\displaylimits_0^{I_{\mathrm{sw}}}P(I)\,\mathrm{d}I\right]
\end{equation}
describes the related probability of the Josephson junction to escape from the zero-voltage state at a nominal switching current $I_{\mathrm{sw}}$ within the interval $\mathrm{d}I$ when a bias current $I$ is injected with a sweep rate $\mathrm{d}I/\mathrm{d}t$. The temperature dependent escape rate
\begin{equation}\label{eq3}
    \tau^{-1} = a_\mathrm{th}\frac{\omega_0}{2\pi}\,\mathrm{e}^{-E_0/k_{\mathrm{B}}T}
\end{equation}
is a function of a temperature and damping dependent thermal prefactor $a_{\mathrm{th}}$, the oscillation frequency $\omega_0~=~\omega_{\mathrm{p}}\left[1-\left(I/I_{\mathrm{c}}\right)^2\right]^{1/4}$ of the Josephson junction with $\omega_{\mathrm{p}}~=~\sqrt{2\pi I_{\mathrm{c}}/\Phi_0C_{\mathrm{JJ}}}$ denoting the plasma frequency and the height of the potential barrier $E_0$ \cite{Wallraff2003}. It can be calculated from the measured switching current distribution $P(I_{\mathrm{sw}})$ of a junction to determine its true critical current $I_{\mathrm{c}}$ by using iterative numerical methods \cite{Castellano1996}. Figure~\ref{fig3} shows as an example of the measured switching current distribution $P(I_{\mathrm{sw}})$ at $T=\SI{4.2}{\kelvin}$ and a fit according to equations~\ref{eq2} and \ref{eq3} for one of our cross-type junctions with a critical current of $I_{\mathrm{c}}= \SI{38.6}{\micro\ampere}$. The dimensionless factor $\kappa = I_\mathrm{c} / I_\mathrm{gap}$, used to calculate the critical current from the measured gap current, is independent of the junction size and constant for an entire junction batch. It was determined by measuring and evaluating the switching current distribution of some representative junctions from each batch.

In order to investigate the spatial profile of the critical current density $j_\mathrm{c}$ along the tunnel barrier, the dependence of the maximum supercurrent $I_{\mathrm{s,max}}$ through the junction on an external magnetic field $B_y$ was measured \cite{Barone1982}. For these measurements, the mu-metal shield of our junction characterization set-up was removed and a Helmholtz coil was attached to the sample holder such that the junction was located in the center of the coil. To analyze the measured data (see, for example, figure~\ref{fig10} in section~\ref{sec:4}), a model of the distribution of the critical current density $j_{\mathrm{c}}(z)$ was generated and the absolute value of the Fourier transform of this model was compared to the measured data.

\begin{figure}
\centering
    \includegraphics[width=0.7\columnwidth]{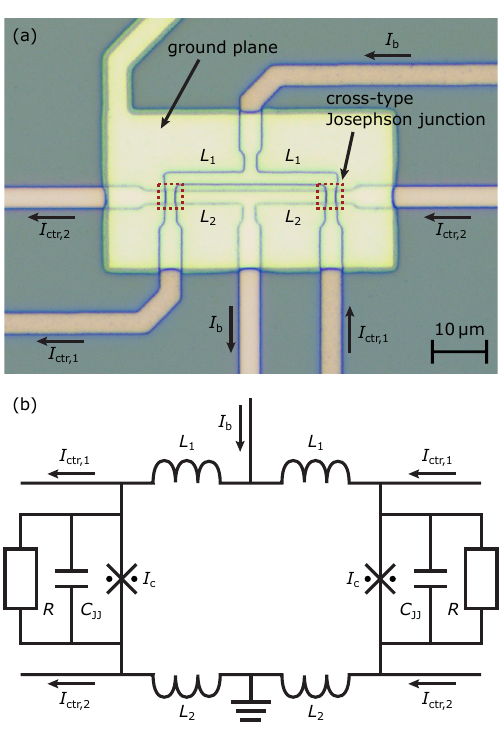}
    \caption{(a) Micrograph and (b) equivalent circuit model of a symmetric, unshunted dc-SQUID based on \qtyproduct{2 x 2}{\micro\meter}-sized cross-type junctions (framed by dashed red boxes) biased by a current $I_{\mathrm{b}}$. The SQUID was designed and fabricated to determine the specific capacitance of our Josephson junctions.}
    \label{fig4}
\end{figure}

The specific capacitance $C_{\mathrm{int}}'$ of our Josephson junctions was determined by observing Shapiro steps in unshunted dc-SQUIDs \cite{Gueret1979, Magerlein1981}. For this, two types of symmetric, unshunted dc-SQUIDs based on cross-type junctions were designed and fabricated. Both variants differ by the junction size (see below). Figure~\ref{fig4} shows a micrograph and the corresponding equivalent circuit model of such a current-biased SQUID comprising \qtyproduct{2 x 2}{\micro\meter}-sized cross-type Josephson junctions. The SQUID loop with inductance $L_{\mathrm{s}}=2(L_1+L_2)$ is composed of two sections. The upper section with inductance $2L_1$ is patterned from the \SI{200}{\nano\meter} thick Nb wiring layer. In contrast, the lower section with inductance $2L_2$ is formed by the \SI{100}{\nano\meter} thick lower Nb layer of the trilayer stack. Both sections are connected via the Josephson junctions. Moreover, feedlines for injecting control currents $I_{\mathrm{ctr,1}}$ and $I_{\mathrm{ctr,2}}$ are connected to both loop sections at the location of the Josephson junctions. A \SI{400}{\nano\meter} thick Nb  ground plane, separated by an insulating SiO$_2$ layer, was patterned on top of all devices to reduce cross-talk between both loop sections \cite{Williams2017}. The resulting parasitic capacitance which is connected in parallel to the capacitance of the two Josephson junctions was estimated to be about \SI{6}{\femto\farad} and corresponds to only \SI{3}{\percent} of the smallest measured capacitance.

To observe the actual resonance steps in the $IV$-characteristic, the maximum supercurrent of a respective sample SQUID was suppressed by applying a control current through one of the feedlines. For $\Phi_{\mathrm{s}} = 2L_{i} I_{\mathrm{ctr,}i} = (n+1/2)\Phi_0$ with $i\in\{1,2\}$, $n \in \mathbb{Z}$ and $\Phi_\mathrm{s}$ denoting the magnetic flux threading the SQUID loop, the maximum supercurrent is at its minimum. At the same time, the periodicity $|\Delta I_{\mathrm{ctr,}i}|~=~\Phi_0/(2L_{\mathrm{i}})$ of the maximum supercurrent modulation with $\Delta I_{\mathrm{ctr,}i}$ being the current difference between two neighboring minima was used to determine the inductance of the SQUID loop \cite{Henkels1978}. The value of $L_{\mathrm{s}}$ for our unshunted SQUIDs with \qtyproduct{2 x 2}{\micro\meter}-sized junctions was simulated to be $L_\mathrm{s} = \SI{14.7}{\pico\henry}$ using InductEx (numeric simulation software by SUN Magnetics (Pty) Ltd.) and is in perfect agreement with the experimental value of $L_\mathrm{s} = \SI{14.0}{\pico\henry}$ taking into account possible fabrication induced size and alignment variances. We also fabricated devices with \qtyproduct{4 x 4}{\micro\meter}-sized cross-type junctions to determine the specific capacitance $C_{\mathrm{int}}'$ for junctions with critical current densities $j_\mathrm{c}~<~\SI{100}{\ampere/\centi\meter\squared}$. Here, the calculated loop inductance is $L_\mathrm{s} = \SI{9.6}{\pico\henry}$ which is again in good agreement with the experimental value of $L_\mathrm{s} = \SI{10.6}{\pico\henry}$.

% --------------------------------------------

\section{Results and discussion}\label{sec:4}

\subsection{Sidewall insulation and characteristic resistance ratio}\label{ssec:4.1}

A key factor for the reliable and reproducible fabrication of high-quality cross-type Josephson junctions based on a Nb/Al-AlO$_x$/Nb trilayer stack is a sufficient galvanic isolation between the base electrode of the junction and a subsequent wiring layer to its top electrode. In our process, this insulation is realized by the quasi-planarizing insulation layer with sufficient thickness to compensate for trenching effects and, even more important, the usage of wet-chemical etching during trilayer etching (see figure~\ref{fig2}(b)). During the wet etching process for removing the Al-AlO$_x$ layer, nitric acid oxidizes the Al surface, while phosphoric acid dissolves the native as well as the continuously formed aluminum oxide. Since niobium, similar to aluminum, oxidizes in nitric acid, but niobium oxide does not dissolve in phosphoric acid \cite{Koch2017}, an oxide layer of a few nm thickness forms on the exposed sidewalls of the patterned Nb/Al-AlO$_x$ stripe and on the surface of the still unstructured lower Nb of the trilayer. This oxide layer serves as a passivation layer. Moreover, compared to plasma induced ion milling no etching residues from redeposited Al atoms \cite{Lehmann1977} appear during wet etching. These residues potentially adhere to the sidewalls of the etched structure forming shorts across the tunnel barrier. Similarly, the passivation layer protects against the formation of shorts originating from potential redeposits during Nb base electrode etching. Overall, the passivation layer formed during wet etching takes on the same task as wet-chemical anodization, however, without the need for a galvanic contact between all patterned trilayer structures.

To prove that wet etching of the Al-AlO$_x$ layer using our acidic etching solution in fact substitutes the anodization of the sidewalls of the patterned trilayer stack, we prepared two distinct batches of cross-type junctions. The Nb/Al-AlO$_x$/Nb trilayer of both batches was sputtered in the same deposition run by placing both substrates side-by-side on the sample holder in the sputter system. For one batch, the Al-AlO$_x$ layer was wet-chemically etched, for the other batch Ar ion milling within the ICP-RIE system was used. For about one half of the Josephson junctions of each batch, the sidewalls were additionally anodized after etching the trilayer stripe.

\begin{figure}
\centering
    \includegraphics[width=0.7\columnwidth]{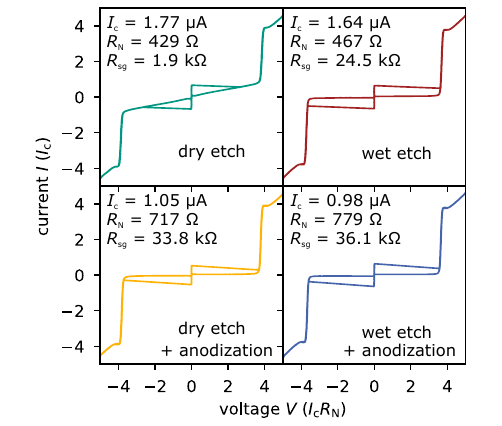}
    \caption{Normalized $IV$-characteristics of anodized (lower row) and non-anodized (upper row) cross-type junctions with a target area of $A_{\mathrm{tar}} = \qtyproduct{2.9 x 2.9}{\micro\meter}$ measured at $ T = \SI{4.2}{\kelvin}$. The Al-AlO$_x$ layer was etched using Ar ion milling in an ICP-RIE system (left column) or using an acidic etching solution based on nitric and phosphoric acid (right column). Note that the current drops to a value below the critical current $I_{\mathrm{c}}$ as the junction jumps into the voltage state due to the voltage dependent junction resistance that is connected in series with the bias resistors (see description of experimental setup in section~\ref{sec:3}).}
    \label{fig5}
\end{figure}

Figure~\ref{fig5} shows current-voltage characteristics of representative cross-type Josephson junctions for each variant. Irrespective of the actual etching technique, the junctions with anodized sidewalls are of high quality which is indicated by very low subgap leakage. Non-anodized junctions for which the Al-AlO$_x$ was wet-chemically etched are of the same quality and have low subgap leakage. In contrast, the $IV$-characteristics of non-anodized, dry etched cross-type junctions exhibit severe leakage. We attribute this to vertical shorts across the tunnel barrier caused by non-passivated Al and Nb redeposits. For the dry etched junctions, where anodization is subsequently performed, these redeposits get oxidized during anodization. We note that the critical current $I_\mathrm{c}$ of anodized junctions is about \SI{40}{\percent} smaller than of non-anodized junctions and attribute this to the reduced junction area due to the anodization induced thick oxide layer on the sidewalls. 

\begin{figure}
\centering
    \includegraphics[width=0.7\columnwidth]{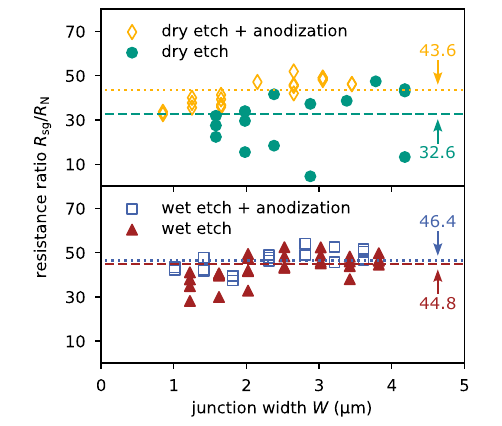}
    \caption{Resistance ratio $R_{\mathrm{sg}}/R_{\mathrm{N}}$ versus junction width for quadratic cross-type junctions based on the same Nb/Al-AlO$_x$/Nb trilayer with anodized (open symbols) and non-anodized (filled symbols) sidewalls, whose Al-AlO$_x$ layer was dry etched (diamonds and circles) or wet etched (squares and triangles). Dotted and dashed horizontal lines mark the mean value for anodized and non-anodized junctions, respectively.}
    \label{fig6}
\end{figure}

The number of redeposits that potentially lead to shorts across the tunnel barrier and hence the subgap leakage are expected to increase with the junction width $W$. Figure~\ref{fig6} displays the dependence of the characteristic resistance ratio $R_{\mathrm{sg}}/R_{\mathrm{N}}$ on the width of the quadratic Josephson junctions from the two examined batches and confirms this hypothesis. We observe an increase of spread with increasing junction size for the dry etched junctions with non-anodized sidewalls (filled circles). Note that the yield of junctions of this variant is only \SI{48}{\percent}, i.e. about every second junction has very high subgap leakage or shows an Ohmic $IV$-characteristic. The fact that this large spread is not observed for anodized junctions from the same batch (open diamonds) is a clear indication that the leakage originates from the sidewalls and not from the tunnel barrier itself. The significantly larger yield of about $90\,\%$ for all other variants further emphasizes that.

The comparison between the non-anodized, wet-chemically etched junctions (filled triangles), the anodized junctions from the same batch (open squares) as well as the dry etched, anodized junctions (see figure~\ref{fig6}) shows that the use of an acidic etching solution for removing the Al-AlO$_x$ layer indeed replaces wet-chemical anodization of the sidewalls of cross-type Josephson junctions. These three variants show a significantly smaller spread of the characteristic resistance ratio as compared to the one with non-anodized dry etched junctions. The ratio tends to get larger the larger the junction is. We attribute this to edge effects that are not caused by redeposits at the sidewalls. Moreover, a direct comparison of the mean characteristic resistance ratios indicates that wet etched, non-anodized junctions (filled triangles) show generally lower subgap leakage than dry etched, anodized specimens (open diamonds). This favors the usage of our fabrication process as compared to processes relying on dry etching the Al-AlO$_x$ layer and subsequent wet-chemical anodization.

% ---------------------

\subsection{Scalability and uniformity of critical current and normal state resistance}\label{ssec:4.2}

\begin{figure}
\centering
    \includegraphics[width=1\columnwidth]{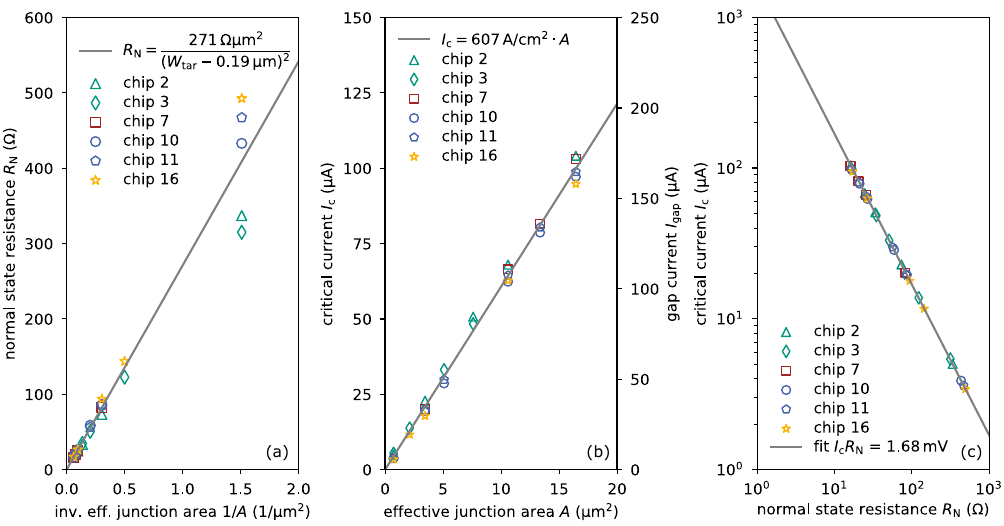}
    \caption{Dependence of (a) the normal state resistance $R_{\mathrm{N}}$ and (b) the critical current $I_{\mathrm{c}}$ on the inverse effective junction area and effective junction area $A=(W_{\mathrm{tar}}+\Delta W)^2$, respectively, for cross-type junctions from different test chips of a single batch internally labeled KA-CJJ-3w9 produced with our anodization-free fabrication process. (a) The solid lines indicate the result of a fit (a) $R_\mathrm{N} = \rho_{\mathrm{N}}/(W_{\mathrm{tar}}+\Delta W)^2$ and (b) $I_\mathrm{c} = j_\mathrm{c} \cdot A$ of the expected linear dependencies and allow to determine the critical current of this batch to be $j_{\mathrm{c}}=\SI{607}{\ampere/\centi\meter\squared}$ and a deviation of the junction size from the target value by only $\Delta W = \SI{-0.19}{\micro\meter}$. (c) Critical current $I_{\mathrm{c}}$ versus normal resistance $R_{\mathrm{N}}$ for junctions from example batch KA-CJJ-3w9. The solid line represents the result of a fit according to $I_{\mathrm{c}}R_{\mathrm{N}} = \mathrm{const}.$ and allows to determine that the $I_{\mathrm{c}}R_{\mathrm{N}}$-product takes a value of \SI{1.68}{\milli\volt}.}
    \label{fig7-8-9}
\end{figure}

For all batches of Josephson junctions that were produced using our anodization-free fabrication process, we checked for the scalability of the critical current $I_\mathrm{c}$ and the normal state resistance $R_\mathrm{N}$ with respect to the junction area as well as for the uniformity of these parameters within a batch. Figure~\ref{fig7-8-9} (a) and (b) show as an example the results for the batch with internal labeling KA-CJJ-3w9 with a critical current density of about \SI{600}{\ampere/\centi\meter\squared}. This batch contains a total of 16 chips with 8 junctions each, 6 of which were measured per cool-down. As expected, both, the critical current $I_\mathrm{c}$ and the normal state resistance $R_{\mathrm{N}}$, scale linearly with the effective and the inverse effective junction area, respectively. We note that the effective junction area $A$ deviates from the target value $A_\mathrm{tar}$ as the lateral junction size deviates by a length $\Delta W$. We determine the deviation by a linear fit $R_\mathrm{N} = \rho_{\mathrm{N}}/(W_{\mathrm{tar}}+\Delta W)^2$, $\rho_\mathrm{N}$ denoting the normal state resistivity, to the data in figure~\ref{fig7-8-9} (a) and find that the size of the cross-type junctions of the batch internally labeled KA-CJJ-3w9 is on average only \SI{0.19}{\micro\meter} smaller than the target value. We attribute this deviation to size variations in the photoresist masks and to a parasitic lateral material loss during both ICP-RIE and isotropic wet etching. Deviations from the linear fit only occur for the smallest junctions with a target area of $A_{\mathrm{tar}} = \qtyproduct{1 x 1}{\micro\meter}$, for which a potentially location dependent variation of the junction size has the greatest effect.

\renewcommand{\arraystretch}{1,3}
\begin{table}
\caption{Summary of the characteristic figures of merit of cross-type junctions from the batch internally labeled KA-CJJ-3w9 measured and determined at \SI{4.2}{\kelvin}. Values in one row correspond to the mean values per junction size across all measured 6 test chips.}
\centering
\vspace{5pt}
\begin{tabular}{ccccccc}
	\hline \\[-1.3em] 
	target size (\SI{}{\micro\meter^2}) & $I_{\mathrm{gap}}$ (\SI{}{\micro\ampere}) & $I_{\mathrm{c}}$ (\SI{}{\micro\ampere}) & $R_{\mathrm{N}}$ ($\mathrm{\Omega}$) & $V_{\mathrm{gap}}$ (mV) & $I_{\mathrm{c}}R_{\mathrm{N}}$ (mV) & $R_{\mathrm{sg}}/R_{\mathrm{N}}$\\[0.35em]
	\hline
	$1\times 1$ & 7.10 & 4.26 & 409 & 2.85 & 1.74 & 47.6\\
	$1.6\times 1.6$ & 21.2 & 12.7 & 133 & 2.84 & 1.69 & 47.1\\
	$2\times 2$ & 33.3 & 20.0 & 84.3 & 2.84 & 1.69 & 46.1\\
    $2.4\times 2.4$ & 50.8 & 30.5 & 55.3 & 2.84 & 1.69 & 47.8\\
    $2.9\times 2.9$ & 82.7 & 49.6 & 34.2 & 2.85 & 1.70 & 42.9\\
    $3.4\times 3.4$ & 108.0 & 64.8 & 25.9 & 2.84 & 1.68 & 37.4\\    $3.8\times 3.8$ & 133.7 & 80.2 & 20.7 & 2.84 & 1.66 & 42.2\\    $4.2\times 4.2$ & 166.0 & 99.6 & 16.5 & 2.84 & 1.64 & 42.7\\
     \hline
\end{tabular}
\label{tab1}
\end{table}

Besides the scalability of the critical current and the normal state resistance, our cross-type junctions also show a high quality that is reflected, for example, by the high average values of the characteristic resistance ratio of $R_{\mathrm{sg}}/R_{\mathrm{N}} > 30$ for small junctions with $W < \SI{2}{\micro\meter}$ and of usually $R_{\mathrm{sg}}/R_{\mathrm{N}} > 40$ for junctions with $W \geq \SI{2}{\micro\meter}$ (compare figure~\ref{fig6}) or the value of the gap voltage $V_{\mathrm{gap}}$. For all batches, $V_\mathrm{gap} > \SI{2.8}{\milli\volt}$, i.e. the gap voltage is very close to the value or the energy gap of Nb \cite{Carbotte1990} and the proximity effect due to the Al layer is negligible. We find $\langle R_{\mathrm{sg}}/R_{\mathrm{N}}\rangle = 44.3$ for batch KA-CJJ-3w9 summarized in table~\ref{tab1} and the values for the gap voltage of all 30 junctions measured are normally distributed with a standard deviation of only $\sigma = \SI{0.01}{\milli\volt}$ from the average $\langle V_\mathrm{gap}\rangle = \SI{2.84}{\milli\volt}$. The $I_{\mathrm{c}}R_{\mathrm{N}}$ product (see figure~\ref{fig7-8-9} (c)) is independent of junction size and is as high as $I_{\mathrm{c}}R_{\mathrm{N}} = \SI{1.7}{\milli\volt}$ for the example batch discussed here. This observation and the almost size independent resistance ratio (see figure~\ref{fig6} and table~\ref{tab1}) indicate that the performance of Nb/Al-AlO$_x$/Nb Josephson tunnel junctions produced with our cross-type fabrication process is not affected by edge effects caused by the process but rather by the intrinsic properties of the tunnel barrier that, of course, could be further optimized.

% ---------------------

\subsection{Profile of the critical current density}\label{ssec:4.3}

\begin{figure}
\centering
    \includegraphics[width=1\columnwidth]{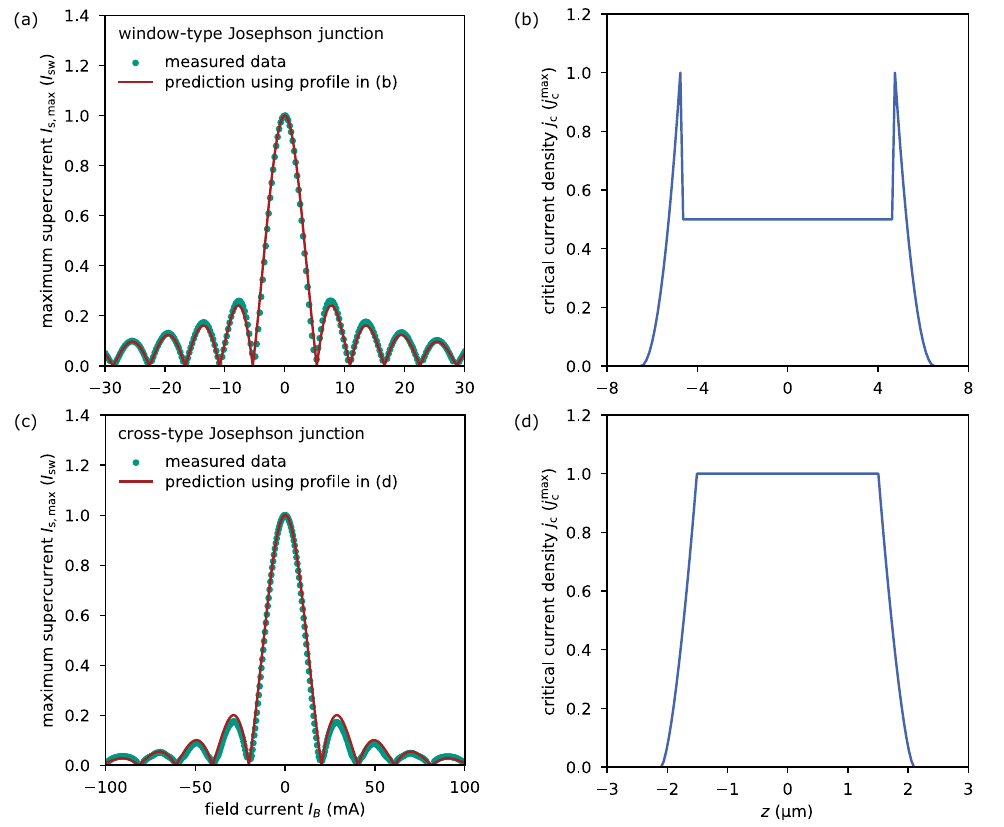}
    \caption{Magnetic field dependence of the normalized maximum supercurrent $I_{\mathrm{s,max}}$ of (a) a \qtyproduct{13 x 13}{\micro\meter} window-type junction and (c) a \qtyproduct{4.2 x 4.2}{\micro\meter} cross-type Josephson junction. For normalization the switching current $I_\mathrm{sw}$ for zero magnetic field was used. The solid red line corresponds to the predicted magnetic field dependence assuming the corresponding critical current density profile $j_{\mathrm{c}}(z)$ plotted in (b) and (d), respectively.}
    \label{fig10}
\end{figure}

The usage of our fabrication process for cross-type Josephson junctions turns out to positively affect the critical current density profile $j_{\mathrm{c}}(z)$ of the tunnel barrier. This can be seen by a comparison between the critical current density distributions shown in figure~\ref{fig10}(b) and figure~\ref{fig10}(d). Both profiles were generated as models to describe the measured dependencies of the maximum supercurrent $I_{\mathrm{s,max}}(I_B)$ of a \qtyproduct{13 x 13}{\micro\meter} window-type and a \qtyproduct{4.2 x 4.2}{\micro\meter} cross-type Josephson junction on the current $I_B$ through the Helmholtz coil of our measurement setup (see section~\ref{sec:3}). The corresponding plots are shown in figure~\ref{fig10}(a) and figure~\ref{fig10}(c), respectively. The batch of Nb/Al-AlO$_x$/Nb window-type junctions was fabricated separately using an anodization-free process based on the one described in \cite{Kempf2013} in which the Al-AlO$_x$ layer was etched by Ar ion milling. The modeled critical current density profile plotted in figure \ref{fig10}(b) is based on two assumptions: 1.) Due to small damages at the edges of the tunnel barrier during dry etching of the Nb top electrode and the Al-AlO$_x$ layer, the flanks of the $j_\mathrm{c}$-profile are not upright but slightly quadratically shaped. 2.) The momentum of the Ar ions during surface cleaning of the top electrode prior to the deposition of the Nb wiring is transferred to the underlying tunnel barrier where it causes damage and thus a reduced critical current density. The area in which this eﬀect occurs is restricted to the size of the window in the insulation layer which is \qtyproduct{11 x 11}{\micro\meter}. The measured magnetic field dependence agrees very well with the prediction from the modeled $j_\mathrm{c}(z)$-profile. The modeled critical current density profile of the cross-type junction shown in figure \ref{fig10}(d) only assumes small damages at the edges of the tunnel barrier during dry etching of the trilayer stack. It clearly exhibits no indentation and yet describes the measured data in figure \ref{fig10}(c) well. Since there is no insulation window on top of the top electrode for a cross-type Josephson junction, the momentum/energy transfer of the Ar ions should be distributed evenly over the entire tunnel barrier, giving rise to a more homogeneous tunnel barrier.

% ---------------------

\subsection{Capacitance measurements}\label{ssec:4.4}

\begin{figure}
\centering
    \includegraphics[width=0.7\columnwidth]{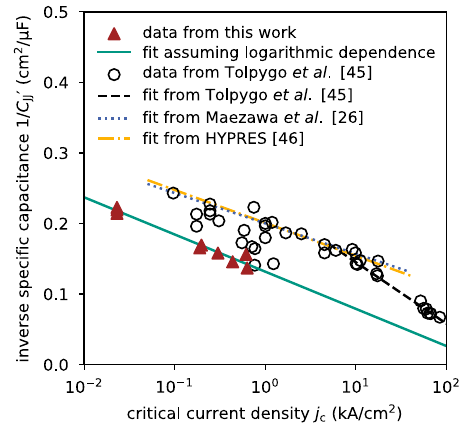}
    \caption{Inverse specific capacitance $C_{\mathrm{JJ}}'$ versus critical current density $j_{\mathrm{c}}$. For our data, each data point represents one characterized unshunted SQUID. The solid green line with $1/C_{\mathrm{JJ}}^\prime = \SI{0.132}{\centi\meter\squared/\micro\farad}-\SI{0.053}{\centi\meter\squared/\micro\farad}\,\log_{10}\left(j_{\mathrm{c}}\,\SI{}{\centi\meter\squared/\kilo\ampere}\right)$ represents the result of a fit to the data assuming a logarithmic dependence. The data points and the dashed line in black are from reference \cite{Tolpygo2017}, the dotted blue line is from reference \cite{Maezawa1995} and the dash-dotted yellow line is from reference \cite{HYPRES} for comparison.}
    \label{fig11}
\end{figure}

\begin{figure}
\centering
    \includegraphics[width=0.7\columnwidth]{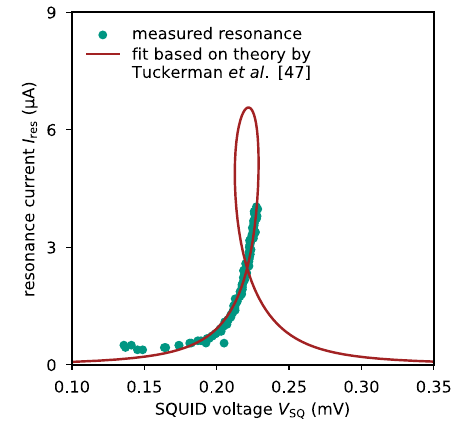}
    \caption{Expected and measured voltage-dependence of the resonance current $I_{\mathrm{res}}$ for an unshunted dc-SQUID with inductance $L_{\mathrm{s}}=\SI{14.0}{\pico\henry}$, resonance voltage $V_{\mathrm{res}}=$ \SI{221}{\micro V}, damping parameter $\Gamma=13.33$ and junction critical current $I_{\mathrm{c}}=$ \SI{9.65}{\micro\ampere}. More details are given in the main text.}
    \label{fig12}
\end{figure}

We measured the capacitance $C_\mathrm{JJ}$ of several cross-type Josephson junctions taken from batches with different critical current densities using unshunted dc-SQUIDs as described in section~\ref{sec:3}. As the parasitic capacitance $C_\mathrm{par}$ is expected to be negligible due to the missing overlap of wiring layers, the measured values should resemble the intrinsic capacitance related to the tunnel barrier. Figure~\ref{fig11} summarizes the result of our measurements and shows the dependence of the inverse junction capacitance per area $C_{\mathrm{JJ}}'$ on the critical current density $j_{\mathrm{c}}$ for each measured SQUID. We note that the specific capacitance $C_\mathrm{JJ}^\prime~=~C_\mathrm{JJ}/A$ is derived from the measured capacitance value $C_\mathrm{JJ}$ as well as the effective junction area $A$ that is determined from fitting the dependence of the normal state resistance on the junction area (see section~\ref{ssec:4.2}). Moreover, figure~\ref{fig12} shows an example of a recorded Shapiro step of a current-biased SQUID with an experimentally determined loop inductance of $L_{\mathrm{s}}~=~\SI{14.0}{\pico\henry}$ and a critical current of $I_{\mathrm{c}}= \SI{9.65}{\micro\ampere}$ of the Josephson junction. The resonance voltage $V_{\mathrm{res}} = \SI{221}{\micro\volt}$ for deriving the junction capacitance
\begin{equation}
    C_\mathrm{JJ} = \frac{\Phi_0^2}{2\pi^2 V_{\mathrm{res}}^2 L_{\mathrm{s}}}
\end{equation}
was determined by fitting the expected shape of the resonance curve to the actual data \cite{Tuckerman1980}. The value of the damping parameter $\Gamma=I_{\mathrm{c}}R/V_{\mathrm{res}}$ required for performing this fit was extracted from its relation to the current ratio $I_{\mathrm{res}}/2I_{\mathrm{c}}$ \cite{Paterno1985}. For the resonance curve shown in figure~\ref{fig12}, $\Gamma = 13.3$. The amplitude of the measured resonance step does not reach the theoretically expected value due to thermal suppression at $T=\SI{4.2}{\kelvin}$.

The solid line in figure~\ref{fig11} represent a fit to the data assuming a logarithmic dependence of the specific capacitance on the critical current density \cite{Maezawa1995,Tolpygo2007}. We find
\begin{equation}
    \frac{1}{C_{\mathrm{JJ}}^\prime} = \qty[per-mode = fraction]{0.132}{\centi\meter\squared\per\micro\farad}
    - \qty[per-mode = fraction]{0.053}{\centi\meter\squared\per\micro\farad} \log_{10}\left(\frac{j_{\mathrm{c}}}{\qty[per-mode = symbol]{}{\kilo\ampere\per\centi\meter\squared}}\right)\,.
\end{equation}

For comparison, we also display the data and derived functional dependencies of the specific capacitance published in \cite{Tolpygo2017} and the references therein. We see that in a critical current density range between approximately \SI{0.01}{\kilo\ampere/\centi\meter\squared} and \SI{10}{\kilo\ampere/\centi\meter\squared} the slope of the functional dependence of our cross-type Josephson junctions is similar to those published by other research groups \cite{Maezawa1995,HYPRES}. Nevertheless, the offset is slightly higher which might be related to different crystal structures of the aluminum oxide barrier.

% --------------------------------------------

\section{Conclusion}
We have developed an anodization-free fabrication process for Nb/Al-AlO$_x$/Nb cross-type junctions. Our process requires only a small number of fabrication steps and can easily be adapted for wafer-scale fabrication. Fabricated junctions are of very high quality as indicated by the measured values for the subgap to normal state resistance ratio and the $I_{\mathrm{c}}R_{\mathrm{N}}$ product. Compared to other junction types manufactured under the same technical conditions, our cross-type Josephson junctions show not only a significantly reduced specific capacitance but also an almost rectangular critical current density profile. Our process hence enables the usage of low capacitance Josephson junctions for superconductor electronic devices such as ultra-low noise dc-SQUIDs, microwave SQUID multiplexers based on non-hysteretic rf-SQUIDs, and RFSQ circuits.

% --------------------------------------------

\section*{Acknowledgments}
We would like to thank A Reifenberger, O Kieler and T Wolf for support during device fabrication and characterization. Moreover, we thank A Ferring, C Jakob, M Schmelz, and R Stolz for fruitful discussions. F Bauer acknowledges former financial support by the Helmholtz Graduate School for Hadron and Ion Research (HGS-HIRe of FAIR). The research leading to these results has also received funding from the European Union’s Horizon 2020 Research and Innovation Programme, under Grant Agreement No. 824109.

% --------------------------------------------

\section*{Data Availability Statement}
The data that support the findings of this study are available from the corresponding author upon reasonable request.

% --------------------------------------------

\section*{References}
\bibliographystyle{iopart-num}
\bibliography{bib.bib}

\providecommand{\newblock}{}
\begin{thebibliography}{10}
\expandafter\ifx\csname url\endcsname\relax
  \def\url#1{{\tt #1}}\fi
\expandafter\ifx\csname urlprefix\endcsname\relax\def\urlprefix{URL }\fi
\providecommand{\eprint}[2][]{\url{#2}}

\bibitem{Clarke2008}
Clarke J and Wilhelm F~K 2008 {\em Nature\/} {\bf 453} 1031--42

\bibitem{Fagaly2006}
Fagaly R~L 2006 {\em Review of Scientific Instruments\/} {\bf 77}

\bibitem{Yohannes2015}
Yohannes D~T, Hunt R~T, Vivalda J~A, Amparo D, Cohen A, Vernik I~V and
  Kirichenko A~F 2015 {\em IEEE Transactions on Applied Superconductivity\/}
  {\bf 25}(3) 1100405

\bibitem{Kunert2017}
Kunert J, Ijsselsteijn R, Il'ichev E, Brandel O, Oelsner G, Anders S, Schultze
  V, Stolz R and Meyer H~G 2017 {\em Low Temperature Physics\/} {\bf 43}(7) 785

\bibitem{Hamilton2000}
Hamilton C~A 2000 {\em Review of Scientific Instruments\/} {\bf 71}

\bibitem{Bluethner1996}
Blüthner K, Götz M, Krech W, Mühlig H, Wagner T, Fuchs H~J, Bluthner K, Gotz
  M, Muhlig H, Schelle D, Fritzsch L, Nachtmann B and Nowack A 1996 {\em
  Journal de Physique IV Proceedings\/} {\bf 111}(C3) 3--163

\bibitem{Dolata2005}
Dolata R, Scherer H, Zorin A~B and Niemeyer J 2005 {\em Journal of Applied
  Physics\/} {\bf 97}(5)

\bibitem{Bhat1999}
Bhat A, Meng X, Whiteley S, Jeffery M and Duzer T~V 1999 {\em IEEE Transactions
  on Applied Superconductivity\/} {\bf 9}(2) 3232--3235

\bibitem{Aumentado2020}
Aumentado J 2020 {\em IEEE Microwave Magazine\/} {\bf 21}(8) 45--59

\bibitem{Rothermel1994}
Rothermel H, Gundlach K~H and Voss M 1994 {\em Le Journal de Physique IV\/}
  {\bf 04}(C6) 267

\bibitem{Westig2012}
Westig M~P, Justen M, Jacobs K, Stutzki J, Schultz M, Schomacker F and Honingh
  N 2012 {\em Journal of Applied Physics\/} {\bf 112}(9) 093919

\bibitem{Tolpygo2019}
Tolpygo S~K, Bolkhovsky V, Rastogi R, Zarr S, Day A~L, Golden E, Weir T~J, Wynn
  A and Johnson L~M 2019 {\em IEEE Transactions on Applied Superconductivity\/}
  {\bf 29}(5) 1101208 ISSN 15582515

\bibitem{Wan2021}
Wan D, Couet S, Piao X, Souriau L, Canvel Y, Tsvetanova D, Vangoidsenhoven D,
  Thiam A, Pacco A, Potočnik A, Mongillo M, Ivanov T, Jussot J, Verjauw J,
  Acharya R, Lazzarino F, Govoreanu B and Radu I~P 2021 {\em Japanese Journal
  of Applied Physics\/} {\bf 60}(SBBI04) ISSN 13474065

\bibitem{Chen2023}
Chen J, Wang Z, Xu D, Qiao H, Li J, Zhong Q, Wang S, Zeng J, Cai J, Zhang M,
  Wang Y, Li X, Zhong Y, Cao W and Wang X 2023 {\em Superconductor Science and
  Technology\/} {\bf 36}(105003) ISSN 13616668

\bibitem{Clarke2004}
Clarke J and Braginski A~I (eds) 2004 {\em The SQUID Handbook: Vol. 1
  Fundamentals and Technology of SQUIDs and SQUID Systems\/} (Wiley-VCH Verlag
  GmbH \& Co. KGaA) ISBN 3527402292

\bibitem{Schmelz2011}
Schmelz M, Stolz R, Zakosarenko V, Anders S, Fritzsch L, Schubert M and Meyer
  H~G 2011 {\em Superconductor Science and Technology\/} {\bf 24}(1) 015005
  ISSN 0953-2048

\bibitem{Gurvitch1983}
Gurvitch M, Washington M~A and Huggins H~A 1983 {\em Applied Physics Letters\/}
  {\bf 42}(5) 472

\bibitem{Dolata1999}
Dolata R, Weimann T, Scherer H~J and Niemeyer J 1999 {\em IEEE Transactions on
  Applied Superconductivity\/} {\bf 9}(2) 3255

\bibitem{Ketchen1991}
Ketchen M~B, Pearson D, Kleinsasser A~W, Hu C~K, Smyth M, Logan J, Stawiasz K,
  Baran E, Jaso M, Ross T, Petrillo K, Manny M, Basavaiah S, Brodsky S, Kaplan
  S~B, Gallagher W~J and Bhushan M 1991 {\em Applied Physics Letters\/} {\bf
  59}(20) 2609--2611

\bibitem{Bao1995}
Bao Z, Bhushan M, Han S and Lukens J~E 1995 {\em IEEE Transactions on Applied
  Superconductivity\/} {\bf 5}(2) 2731--2734

\bibitem{Watanabe2004}
Watanabe M, Nakamura Y and Tsai J~S 2004 {\em Applied Physics Letters\/} {\bf
  84}(3) 410--412

\bibitem{Harada1994}
Harada Y, Haviland D~B, Delsing P, Chen C~D and Claeson T 1994 {\em Applied
  Physics Letters\/} {\bf 65}(5) 636--638

\bibitem{Dang1991}
Dang H and Radparvar M 1991 {\em IEEE Transactions on Magnetics\/} {\bf 27}(2)
  3157

\bibitem{Anders2009}
Anders S, Schmelz M, Fritzsch L, Stolz R, Zakosarenko V, Schönau T and Meyer
  H~G 2009 {\em Superconductor Science and Technology\/} {\bf 22}(6) 064012

\bibitem{Kaiser2011}
Kaiser C, Meckbach J~M, Ilin K~S, Lisenfeld J, Schäfer R, Ustinov A~V and
  Siegel M 2011 {\em Superconductor Science and Technology\/} {\bf 24}

\bibitem{Maezawa1995}
Maezawa M, Aoyagi M, Nakagawa H, Kurosawa I, Takada S and Tuckerman A 1995 {\em
  Applied Physics Letters\/} {\bf 66}(16) 2134

\bibitem{Tolpygo2007}
Tolpygo S~K, Yohannes D, Hunt R~T, Vivalda J~A, Donnelly D, Amparo D and
  Kirichenko A~F 2007 {\em IEEE Transactions on Applied Superconductivity\/}
  {\bf 17}(2) 946

\bibitem{Kuroda1988}
Kuroda K and Yuda M 1988 {\em Journal of Applied Physics\/} {\bf 63}(7) 2352

\bibitem{Du2007}
Du J, Charles A~D~M, Petersson K~D and Preston E~W 2007 {\em Superconductor
  Science and Technology\/} {\bf 20}(11) S350

\bibitem{Likharev1986}
Likharev K~K 1986 {\em Dynamics of Josephson Junctions and Circuits\/} (Taylor
  \& Francis) ISBN 9782881240423

\bibitem{Likharev1979}
Likharev K~K 1979 {\em Reviews of Modern Physics\/} {\bf 51}(1) 101

\bibitem{Fulton1974}
Fulton T~A and Dunkleberger L~N 1974 {\em Physical Review B\/} {\bf 9}(11) 4760

\bibitem{Danchi1984}
Danchi W~C, Hansen J~B, Octavio M, Habbal F and Tinkham M 1984 {\em Physical
  Review B\/} {\bf 30} 2503--2516

\bibitem{Wallraff2003}
Wallraff A, Lukashenko A, Coqui C, Kemp A, Duty T and Ustinov A~V 2003 {\em
  Review of Scientific Instruments\/} {\bf 74}(8) 3740

\bibitem{Castellano1996}
Castellano M~G, Leoni R, Torrioli G, Chiarello F, Cosmelli C, Costantini A,
  Diambrini-Palazzi G, Carelli P, Cristiano R and Frunzio L 1996 {\em Journal
  of Applied Physics\/} {\bf 80}(5) 2922

\bibitem{Barone1982}
Barone A and Paternò G 1982 {\em Physics and Applications of the Josephson
  Effect\/} (John Wiley \& Sons, Inc.) ISBN 9783527602780

\bibitem{Gueret1979}
Guéret P 1979 {\em Applied Physics Letters\/} {\bf 35}(11) 889

\bibitem{Magerlein1981}
Magerlein J~H 1981 {\em IEEE Transactions on Magnetics\/} {\bf MAG-17}(1) 286

\bibitem{Williams2017}
Williams T 2017 {\em EMC for Product Designers\/} 5th ed (Newnes) ISBN
  978-0-08-101016-7

\bibitem{Henkels1978}
Henkels W~H 1978 {\em Applied Physics Letters\/} {\bf 32}(12) 829

\bibitem{Koch2017}
Koch C and Rinke T~J 2017 {\em Photolithography: Basics of Microstructuring\/}
  1st ed (MicroChemicals GmbH) ISBN 978-3-9818782-1-9

\bibitem{Lehmann1977}
Lehmann H~W, Krausbauer L and Widmer R 1977 {\em Journal of Vacuum Science and
  Technology\/} {\bf 14}(1) 281

\bibitem{Carbotte1990}
Carbotte J~P 1990 {\em Reviews of Modern Physics\/} {\bf 62}(4) 1027--1157

\bibitem{Kempf2013}
Kempf S, Ferring A, Fleischmann A, Gastaldo L and Enss C 2013 {\em
  Superconductor Science and Technology\/} {\bf 26}(6) 065012

\bibitem{Tolpygo2017}
Tolpygo S~K, Bolkhovsky V, Zarr S, Weir T~J, Wynn A, Day A~L, Johnson L~M and
  Gouker M~A 2017 {\em IEEE Transactions on Applied Superconductivity\/} {\bf
  27}(4) 1100815

\bibitem{HYPRES}
Hypres niobium integrated circuit fabrication, process \#03-10-45, design
  rules, revision \#25, 12/12/2012
  \urlprefix\url{https://www.hypres.com/wp-content/uploads/2010/11/DesignRules-4.pdf}

\bibitem{Tuckerman1980}
Tuckerman D~B and Magerlein J~H 1980 {\em Applied Physics Letters\/} {\bf
  37}(2) 241

\bibitem{Paterno1985}
Paternò G, Cucolo A~M and Modestino G 1985 {\em Journal of Applied Physics\/}
  {\bf 57}(5) 1680

\end{thebibliography}

\end{document}